\documentclass[floatfix,aps,prd,reprint,nofootinbib,superscriptaddress,amsmath,amssymb]{revtex4-2}

\usepackage[italicdiff]{physics}
\usepackage{blindtext}
\usepackage{graphicx}
\usepackage{dcolumn}
\usepackage{hyperref}
\usepackage{bm}
\usepackage{xcolor}
\usepackage[normalem]{ulem}

\def\eq{Eq.}
\def\eqs{Eqs.}
\def\A{\mathcal A}
\def\I{\mathcal I}
\def\J{\mathcal J}
\def\K{\mathcal K}
\def\E{\mathcal E}
\def\B{\mathcal B}
\def\C{\mathcal C}
\def\AE{A\&E}

\begin{document}

\title{Critical phenomena in the collapse of quadrupolar and hexadecapolar gravitational waves}

\author{Thomas W. Baumgarte}
\email{tbaumgar@bowdoin.edu}
\affiliation{Department of Physics and Astronomy, Bowdoin College, Brunswick, Maine 04011, USA}

\author{Carsten Gundlach}
\email{c.j.gundlach@soton.ac.uk}
\affiliation{School of Mathematical Sciences, University of Southampton,
  Southampton SO17 1BJ, United Kingdom} 

\author{David Hilditch}
\email{david.hilditch@tecnico.ulisboa.pt}
\affiliation{
  Centro de Astrof\'{\i}sica e Gravita\c c\~ao -- CENTRA,
  Departamento de F\'{\i}sica, Instituto Superior T\'ecnico -- IST,
  Universidade de Lisboa -- UL, Av.\ Rovisco Pais 1, 1049-001 Lisboa,
  Portugal}

\begin{abstract}
We report on numerical simulations of critical phenomena near the threshold of black hole formation in the collapse of axisymmetric gravitational waves in vacuum.  We discuss several new features of our numerical treatment, and then compare results obtained from families of quadrupolar and hexadecapolar initial data.   Specifically, we construct (nonlinear) initial data from quadrupolar and hexadecapolar, time-symmetric wavelike solutions to the linearized Einstein equations (often referred to as Teukolsky waves), and evolve these using a shock-avoiding slicing condition.   While our degree of fine-tuning to the onset of black-hole formation is rather modest, we identify several features of the threshold solutions formed for the two families.   Both threshold solutions appear to display an at least approximate discrete self-similarity with an accumulation event at the center, and the characteristics of the threshold solution for the quadrupolar data are consistent with those found previously by other authors.  The hexadecapolar threshold solution appears to be distinct from the quadrupolar one, providing further support to the notion that there is no universal  critical solution for the collapse of vacuum gravitational waves.
\end{abstract}
\maketitle


\section{Introduction}
\label{sec:intro}

Critical phenomena in gravitational collapse, first reported by Choptuik in his seminal paper \cite{Cho93}, refer to properties of solutions to Einstein's equations close to the onset of black-hole formation.  Choptuik performed numerical evolution calculations for a massless scalar field, minimally coupled to Einstein's equations, in spherical symmetry.  Considering different families of initial data, parameterized by $p$, say, he noted the existence of a critical parameter $p_*$ that separates {\em supercritical} data, which lead to the formation of a black hole, from {\em subcritical} data, which leave behind flat space after the wave disperses.  Critical phenomena, with intriguing resemblance to similar phenomena in other fields of physics, then emerge in the vicinity of $p_*$.

Specifically, Choptuik observed that, for initial data fine-tuned to $p_*$, the solution approaches a {\em critical solution} that contracts {\em self-similarly}.  Moreover, for supercritical data close to $p_*$, the mass $M$ of black holes displays {\em power-law scaling}
\begin{equation} \label{mass_scaling}
    M \simeq (p - p_*)^\gamma, 
\end{equation}
where the {\em critical exponent} $\gamma$ is {\em universal} in the sense that it does not depend on the family of initial data.

Triggered by Choptuik's discovery, a number of different groups and researchers have studied critical collapse for different matter models, symmetries, number of dimensions, and asymptotics (see, e.g., \cite{GunM07} for a review and references).   At least in spherical symmetry, the phenomena observed by Choptuik can be understood in terms of a self-similar critical solution that is {\em universal} for all families of initial data within a given matter model, and that possesses exactly one unstable mode whose growth rate is described by a Lyapunov exponent $\lambda$.   Any quantity  with dimension of  mass (or length) resulting from the dynamical evolution is then described by a scaling law (\ref{mass_scaling}) with the critical exponent given by the inverse of the Lyapunov exponent, $\gamma = 1/ \lambda$ \cite{KoiHA95,Mai96}.   Accordingly, $\gamma$ is unique for a given matter model, and scaling relations similar to (\ref{mass_scaling}) apply to both supercritical (e.g.~the black-hole mass) and subcritical data (e.g.~quantities formed from the maximum attained spacetime curvature, see \cite{GarD98}).   The critical solution can either be {\em continuously} self-similar, describing a continuous contraction (for example for perfect fluids, see \cite{EvaC94}), or it can be {\em discretely} self-similar (from now on, DSS), describing an oscillation that is superimposed on the contraction (for example for scalar fields).   For a DSS critical solution, scaling laws like (\ref{mass_scaling}) feature a periodic ``wiggle" superimposed on the power-law scaling, whose periodicity is related to that of the self-similar solution (see \cite{Gun97,HodP97}).

Shortly after Choptuik's announcement, Abrahams and Evans reported very similar critical phenomena in the collapse of axisymmetric vacuum gravitational waves (\cite{AbrE93,AbrE94}, hereafter \AE).  Even though the case for an exact DSS in their remarkable simulations was less convincing than in Choptuik's calculations, the authors attributed this to numerical error, which, given the absence of spherical symmetry, was necessarily larger.  Regardless of these issues, the findings of \AE\ seemed to suggest that the characteristics of critical collapse in spherical symmetry -- self-similarity, scaling, and universality -- should similarly apply in the absence of spherical symmetry.

While a number of authors have performed numerical simulations of nonlinear gravitational waves (e.g.~\cite{Alcetal00,GarD01,Rin08,Sor11,Hiletal13}) it has proven difficult to reproduce the results of \AE.  In the meantime, several authors studied critical collapse for non-vacuum spacetimes in the absence of spherical symmetry and observed qualitatively new features.  In the critical collapse of scalar fields in axisymmetry with an additional reflection symmetry through the equatorial plane, for example, it was found that, for sufficiently large departure from spherical symmetry and exquisite fine-tuning, a ``bifurcation" occurs, leading to the formation of two separate centers of collapse away from the center of symmetry (see \cite{ChoHLP03,Bau18}).  Studying the gravitational collapse of dipolar electromagnetic waves, we found that the critical solution is approximately, but not exactly DSS (see \cite{BauGH19}).  Moreover, the authors of \cite{PerB21} found that the critical solution found in the collapse of quadrupolar electromagnetic waves is different from that for dipolar waves, suggesting that the critical solution is not universal. To emphasize this, we will refer to these solutions as {\em threshold} solutions, and will reserve the term {\em critical} solution for cases in which it is universal, i.e.~independent of the family of initial data.  Similar results were found by \cite{SuaVH21} for critical collapse in an analytical model problem.  All the above suggests that, in the absence of spherical symmetry, critical phenomena are not characterized by a universal, exactly self-similar critical solution.

Significant progress in numerical simulations of the collapse of gravitational waves has recently been reported by (\cite{HilWB17,LedK21,FerRABH22}).   The authors of \cite{LedK21}, in particular, considered different families of initial data, and found that they lead to different critical exponents, suggesting that the corresponding threshold solutions are also distinct.  The authors of all three papers also found that, while the maximum curvature attained in subcritical evolutions satisfies approximate power-law scaling, wiggles superimposed on these power laws are not strictly periodic, indicating that the underlying threshold solutions are not exactly DSS. All of these observations are in accordance with the findings discussed above, and raise the question whether the characteristics of critical phenomena observed in spherical symmetry also apply in the absence of spherical symmetry. 

The purpose of this paper is to complement the results of \cite{HilWB17,LedK21,FerRABH22} with independent simulations of vacuum critical collapse that differ from the above in several ways.  Specifically, we adopt families of initial data that are based on time-symmetric Teukolsky waves, both quadrupolar \cite{Teu82} and hexadecapolar \cite{Rin09}, supplementing the families considered by \cite{HilWB17,LedK21,FerRABH22} (see Sect.~\ref{sec:indata} below).  We also describe significant improvements that resulted from replacing the much more common 1+log slicing condition \cite{BonMSS95} (used, for example, in \cite{BauGH19,PerB21}) with a shock-avoiding slicing condition (see \cite{Alc97,Alc03}; Sect.~\ref{sec:slicing}).  Finally, we use both scalar invariants of the Weyl tensor, rather than just the Kretschmann tensor, as a diagnostic tool (Sect.~\ref{sec:diagnostics}).  

While the degree of fine-tuning to the threshold parameter that we achieve is more modest than in the calculations of \cite{AbrE93,AbrE94,HilWB17,LedK21,FerRABH22}, our results, discussed in Sect.~\ref{sec:results}, provide independent support for our emerging understanding of critical collapse of gravitational waves (see also \cite{Bauetal23}).  In particular, we provide further evidence for the absence of a universal critical solution, while suggesting there exist families of gravitational-wave initial data for which the threshold solution is at least approximately DSS with a single accumulation point. 

\section{Numerical Methods}
\label{sec:numerics}

All numerical results presented in this paper were obtained with a code that solves the Baumgarte-Shapiro-Shibata-Nakamura (BSSN) formulation \cite{NakOK87,ShiN95,BauS98} of Einstein's equations in spherical polar coordinates.  In particular, our implementation adopts a reference-metric formulation (e.g.~\cite{BonGGN04,ShiUF04,Bro09,Gou12,MonC12}) together with a proper rescaling of all tensor components to handle the coordinate singularities at the origin and on the coordinate axis.  General features of this code are discussed in \cite{BauMCM13}; more recent improvements include the replacement of the partially-implicit Runge-Kutta method with a fourth-order Runge-Kutta method of lines for the time evolution (see \cite{BauGH19}), the implementation of an asymptotically logarithmic radial grid (following the prescription of \cite{RucEB18}), and the ability to regrid the radial grid in order to achieve higher resolution close to the origin later in the evolution (see \cite{BauG16}).  Our current implementation also uses eighth-order finite-difference stencils to evaluate all spatial derivatives.   

Since most features of our code have been discussed elsewhere already, we focus here on aspects and improvements that are relevant for simulations of critical phenomena in the collapse of gravitational waves.

\subsection{Initial Data}
\label{sec:indata}

Two different approaches are commonly adopted to construct initial data describing vacuum gravitational waves.  

In one approach, leading to {\em Brill waves} (see \cite{Bri59}), the spacetime is assumed to be axisymmetric and to admit a moment of time symmetry.  Departures from flatness can then be described in terms of a seed function in such a way that the Hamiltonian constraint reduces to a linear elliptic equation whose solution provides nonlinear initial data at the moment of time symmetry.  Brill data have been adopted in numerous simulations of gravitational waves, including in some recent studies of critical phenomena in their collapse (see \cite{HilWB17,LedK21,FerRABH22}).  

In this paper we adopt an alternative approach that is based on analytical wave solutions to the linearized Einstein equations in transverse-traceless (TT) gauge, often referred to as {\em Teukolsky waves}.  Quadrupolar solutions are presented in \cite{Teu82}, while generalizations for higher multipole moments are provided in \cite{Rin09}.  In this paper we will consider both quadrupolar ($\ell = 2$) and hexadecapolar ($\ell = 4$) waves.  

Like Brill waves, Teukolsky waves are constructed from a seed function $F$ which, in this case, can depend on the combinations $r + t$ or $r - t$ describing ingoing or outgoing waves.  For even-parity modes, one computes from the seed function $F$ and its derivatives the functions $A_\ell(t,r)$, $B_\ell(t,r)$, and $C_\ell(t,r)$ (see \eqs~\ref{app:radial_2} and \ref{app:radial_4} in Appendix \ref{sec:appA}), whose products with angular functions (see \eqs~\ref{app:angular_2} and \ref{app:angular_4}) then describe the components of the spatial metric $\gamma_{ij}$ (see \ref{rinne_metric}) in TT gauge.

In order to obtain nonlinear solutions to Einstein's constraint equations we follow the above prescription and construct a linear combination of ingoing and outgoing waves of a given multipole moment $\ell$ so that the instant $t=0$ corresponds to a moment of time symmetry.  Accordingly, the extrinsic curvature $K_{ij}$ vanishes at this moment and the data satisfy the momentum constraints identically.  We then identify the conformally related metric $\bar \gamma_{ij}$ with the linear wave metric as constructed above and solve the Hamiltonian constraint for the conformal factor $\psi$ iteratively in order to reduce its violation by several orders of magnitude.  Given $\psi$, the physical metric $\gamma_{ij} = \psi^4 \bar \gamma_{ij}$ together with $K_{ij} = 0$ then provide nonlinear solutions to the constraint equations describing gravitational-wave initial data.  Because of nonlinear coupling, these solutions no longer represent a single multipole moment, but we will refer to them by the multipole moment of the underlying linear solution regardless.

Specifically, we choose the seed function
\begin{equation} \label{seed}
    F(t,r) = \frac{\A \lambda^{3 + \ell}}{2} \left( u \left( e^{-u_+^2} + e^{-u_-^2} \right) + v \left(e^{-v_+^2} + e^{-v_-^2} \right) \right)
\end{equation}
where $\A$ is the dimensionless amplitude and $\lambda$ parameterizes the wavelength.  We have also defined the dimensionless combinations
\begin{subequations}
\begin{align}
    u \equiv (r - t) / \lambda,~~~~~~~v \equiv (r + t) / \lambda,
\end{align}
as well as 
\begin{align}    
    u_\pm \equiv (r - t \pm r_0) / \lambda,~~~~~~~v_\pm \equiv (r + t \pm r_0)/ \lambda,
\end{align}
\end{subequations}
where $r_0$ parameterizes the location of the resulting wave package at the moment of time symmetry $t = 0$.  Throughout this paper we adopt $\lambda$ as our ``code unit", so that all dimensional results are given in units of $\lambda$.   For all simulations presented in this paper we adopt $r_0 = 2$.

One important improvement in our code over our previous implementations is related to the computation of the functions $A_\ell(t,r)$, $B_\ell(t,r)$, and $C_\ell(t,r)$ from the seed function $F$.  Specifically, these functions, see \eqs~(\ref{app:angular_2}) and (\ref{app:radial_4}) for $\ell = 2$ and $\ell = 4$, involve multiple terms that are divided by high powers of $r$.  For small values of $r$, the individual terms in these functions can evidently become very large.  If the functions were implemented as written, round-off error would lead to imperfect cancellation between the individual terms, and result in large numerical error in the functions $A_\ell(t,r)$, $B_\ell(t,r)$, and $C_\ell(t,r)$ in the vicinity of the origin.  In order to avoid this error, it is important to take advantage of these cancellations analytically.  In our implementation here we use a Taylor expansion up to order $r^6$ (for $\ell = 2$)  or $r^8$ (for $\ell = 4$) about the origin $r = 0$ and match this expansion to a direct implementation at a suitable cut-off radius.\footnote{For $r_0 = 0$ and $t = 0$ the functions $A_\ell(t,r)$, $B_\ell(t,r)$, and $C_\ell(t,r)$ can also be simplified analytically, resulting in expressions that no longer include divisions by $r$; see \eqs~(A3) of \cite{FerBH21} for an example.}

\begin{figure*}[t]
    \centering
    \includegraphics[width = 0.45 \textwidth]{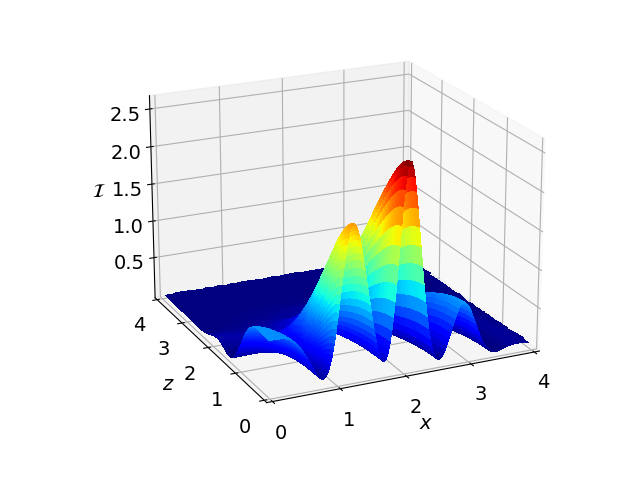}
    \includegraphics[width = 0.45 \textwidth]{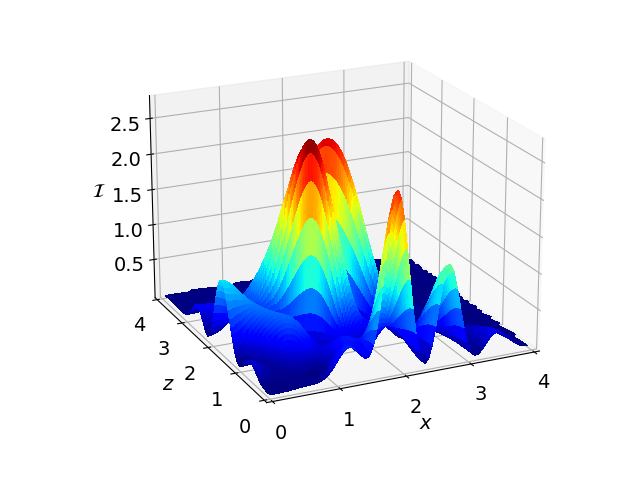}

    \includegraphics[width = 0.45 \textwidth]{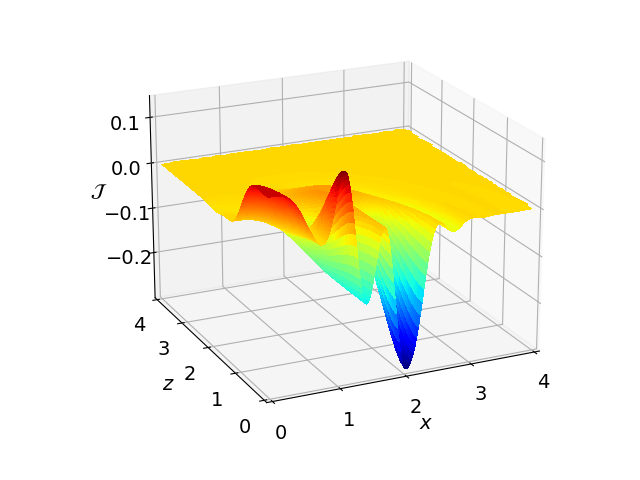} 
    \includegraphics[width = 0.45 \textwidth]{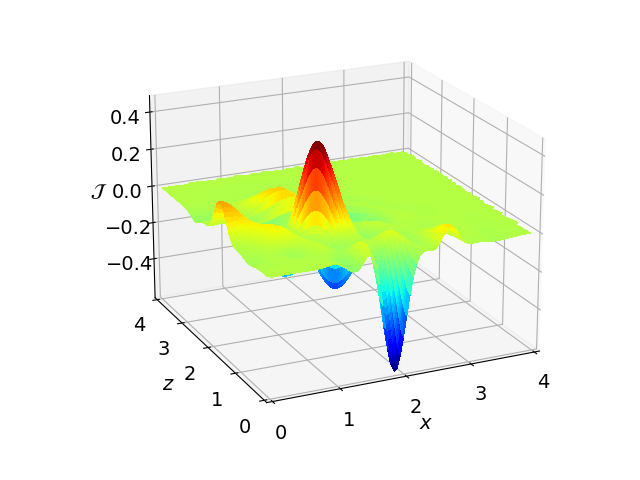} 
    \caption{The curvature invariants $\I$ (top row) and $\J$ (bottom row; see \eqs~(\ref{invariants}) below) at the initial moment of time symmetry $t=0$ for near-critical axisymmetric Teukolsky waves.  In the left column we show quadrupolar ($\ell = 2$) initial data with an amplitude $\A = 0.00495634$, and in the right column hexadecapolar ($\ell = 4$) data with an amplitude of $\A = 0.00004251$.  We construct the Cartesian coordinates $x = r \sin \theta$ and $z = r \cos \theta$ from the spherical polar coordinates adopted in our code in the usual way; in particular, the $z$-axis is the axis of symmetry,  while the $x$-axis lies in the equatorial plane.}
    \label{fig:indata}
\end{figure*}

We also comment on the initialization of the conformal connection functions in the BSSN formalism, often denoted $\bar \Gamma^i \equiv \bar \gamma^{jk} \bar \Gamma^i_{jk} = - \partial_j \bar \gamma^{ij}$ in the context of Cartesian coordinates, or $\bar \Lambda^i = \bar \gamma^{jk} \Delta \Gamma^i_{jk}$ in the context of a reference-metric formalism.  To reduce numerical error, the conformal connection functions are initialized using analytical expressions involving the above functions $A_\ell(t,r)$, $B_\ell(t,r)$, and $C_\ell(t,r)$ (see \eqs~\ref{app:radial_2} and \ref{app:radial_4}), the angular functions (\ref{app:angular_2}) and (\ref{app:angular_4}), as well as their derivatives.  Inside the cut-off radius for the Taylor expansion we evaluate the radial derivatives of $A_\ell(t,r)$, $B_\ell(t,r)$, and $C_\ell(t,r)$ analytically from the power-law expansion, and outside we compute them numerically but with a stencil that is much finer than that used in the evolution code, resulting in smaller errors.  The angular derivatives of the angular functions (\ref{app:angular_2}) and (\ref{app:angular_4}) are easy to compute analytically everywhere.

In this paper we focus on axisymmetric data (i.e.~$m = 0$) and even-order multipoles, for which the solutions are also symmetric across the equator.  The origin $r = 0$ therefore represents the geodesic worldline of a preferred observer in our simulations, and we may restrict our numerical grid to one hemisphere.  We present results for quadrupolar data with $\ell = 2$ in Sect.~\ref{sec:quadru} and for hexadecapolar data with $\ell = 4$ in Sect.~\ref{sec:hexadeca}.  As an illustration of these initial data we show in Fig.~\ref{fig:indata} the curvature invariants $\I$ and $\J$ (see Sect.~\ref{sec:diagnostics}) for both quadrupolar and hexadecapolar, near-critical Teukolsky waves at the initial moment of time symmetry.

\subsection{Slicing condition}
\label{sec:slicing}

During the dynamical evolution of our initial data we impose a {\em Bona-Mass\'o slicing} condition
\begin{equation} \label{bonamasso}
    \alpha n^a \partial_a \alpha = (\partial_t - \beta^i \partial_i) \alpha = - \alpha^2 f(\alpha) K,
\end{equation}
where $\alpha$ is the lapse function, $\beta^i$ the shift vector, $n^a = \alpha^{-1} (1, \beta^i)$ the spacetime normal on spatial slices (of constant coordinate time $t$), $K$ is the mean curvature, and the {\em Bona-Mass\'o function} $f(\alpha)$ is a yet-to-be-determined function of the lapse (see \cite{BonMSS95}).  A very common choice for $f(\alpha)$ is
\begin{equation} \label{1+log}
    f(\alpha) = \frac{2}{\alpha}.
\end{equation}
In the absence of a shift and up to a constant of integration, \eq~(\ref{bonamasso}) can then be integrated to yield $\alpha = 1 + \log (\det(\gamma_{ij}))$, which lends this condition its name {\em 1+log slicing}.  1+log slicing has been used in numerous simulations, including in the first successful BSSN simulations of binary black holes (see \cite{CamLMZ06,BakCCKM06}) as well as many follow-up simulations.  

However, 1+log slicing is also known to develop coordinate shocks in some situations, even in flat spacetimes (see \cite{Alc97,Alc03,Alc05}).  When this happens, the lapse typically develops increasingly steep gradients, and the mean curvature $K$ increasingly large spikes, ultimately leading to the code crashing (see also Fig.~1 of \cite{BauH22} for a recent numerical example).  Similar behavior was reported by \cite{Hiletal13} for simulations of the collapse of gravitational waves with 1+log slicing, where it prevented a study of critical phenomena close to the black-hole threshold.

As an alternative to 1+log slicing, Alcubierre therefore suggested a {\em shock-avoiding slicing} condition with 
\begin{equation} \label{sa}
    f(\alpha) = 1 + \frac{\kappa}{\alpha^2},
\end{equation}
where $\kappa > 0$ is a constant (see \cite{Alc97,Alc03}).  Even though this condition has the unusual property that it allows the lapse to become negative during dynamical simulations (which may explain why it has not been adopted widely), it has recently been shown to perform similarly to 1+log slicing in terms of accuracy and stability for a number of test cases (see \cite{BauH22}).  Analytical results on static black-hole trumpet slices of the Schwarzschild spacetime satisfying the shock-avoiding slicing condition have been presented in \cite{BaudeO22}, and dynamical perturbations of such slices have been explored in \cite{LiBDdeO23}.  Shock-avoiding slicing has also been used in the critical collapse simulations of \cite{JimVA21}.

\begin{figure}
    \centering
    \includegraphics[width = 0.45 \textwidth]{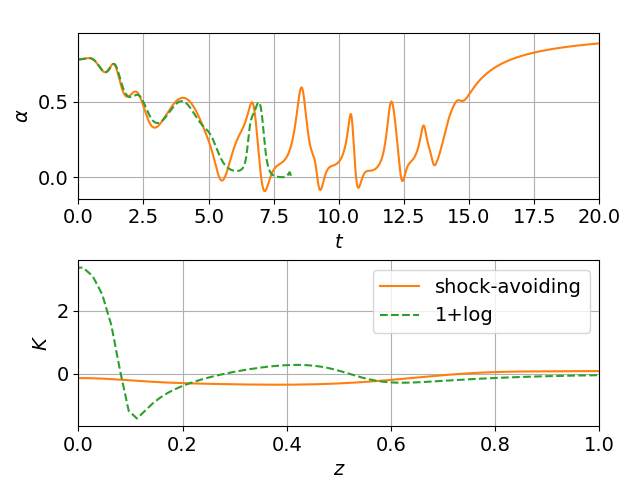}
    \caption{Comparison of the evolution of a near-critical $\ell = 2$ Teukolsky wave (with amplitude $\A = 0.004955$) with 1+log versus shock-avoiding slicing.  In the top panel we show the lapse $\alpha$ at the origin as a function of coordinate time $t$; in the bottom panel we show the mean curvature $K$ along the symmetry axis at coordinate time $t \simeq 7.83$, shortly before the evolution with 1+log slicing crashes.  We note that the shock-avoiding slicing condition results in the lapse becoming negative during short intervals of time.}
    \label{fig:slicing}
\end{figure}

Replacing the 1+log slicing condition has been crucial for our simulations here.  As a demonstration, in Fig.~\ref{fig:slicing} we compare results from the evolution of identical initial data with 1+log and shock-avoiding slices.  Specifically, we evolve a near- but subcritical $\ell = 2$ wave with amplitude $\A = 0.004955$.  In the top panel we show the lapse $\alpha$ at the origin $r = 0$ as a function of coordinate time $t$.  At early times both conditions lead to quite similar behavior, but, while the shock-avoiding slicing allows the lapse to perform multiple oscillations (taking negative values some of the time) before asymptoting to unity as the wave disperses to infinity, the 1+log slicing leads to the code crashing at around $t \simeq 8.2$.  In the bottom panel we show profiles of the mean curvature along the symmetry axis at time $t \simeq 7.83$, shortly before the evolution with 1+log slicing crashes.  For 1+log slicing, the mean curvature shows a spike developing at around $z \simeq 0.1$, which keeps growing and ultimately causes the calculation to fail.  For shock-avoiding slicing, on the other hand, the mean curvature does not develop such a spike, and instead remains smooth. 

While using shock-avoiding slicing rather than 1+log slicing has led to a dramatic improvement in our simulations here, this is not the only option, of course.  Maximal slicing (which was adopted by \cite{AbrE93,AbrE94}) or an approximate maximal slicing condition (see \cite{KhiL18,LedK21}) have also been used successfully in simulations of the collapse of gravitational waves, as have been other gauge conditions imposed via gauge source functions in the context of the generalized harmonic system (see \cite{HilWB17,FerRABH22}).  As an attractive feature of shock-avoiding slicing we point out that it is {\em very} easy to implement, especially in codes that use 1+log slicing already.

\subsection{Diagnostics}
\label{sec:diagnostics}

We diagnose the geometry in our dynamical simulations by evaluating scalar curvature invariants $\I$ and $\J$ of the Weyl tensor $C_{abcd}$, which, for the vacuum spacetimes considered here, is equal to the spacetime Riemann tensor ${}^{(4)}R_{abcd}$.  

We compute the invariants from the electric and magnetic parts of the Weyl tensor (see, e.g., \cite{SteKMHH03}, as well as \cite{BurBB06} for examples).  The electric part $\E_{ij}$ can be computed from
\begin{equation}
    \E_{ij} = R_{ij} + K K_{ij} - K_{ik} K^k_{~j},
\end{equation}
where $R_{ij}$ is the spatial Ricci tensor, $K_{ij}$ the extrinsic curvature, and its trace $K = \gamma_{ij} K^{ij}$ the mean curvature.  The magnetic part $\B_{ij}$ is 
\begin{equation}
    \B_{ij} = \epsilon_{(i|}^{~kl} \nabla_k K_{|j)l},
\end{equation}
where $\epsilon_{ijk}$ is the spatial Levi-Civita tensor, and where the parantheses denote symmetrization of the indices $i$ and $j$.  Both $\E_{ij}$ and $\B_{ij}$ are symmetric and tracefree.   We then form the complex tensor
\begin{equation}
\C^i_{~j} \equiv \E^i_{~j} + i \B^i_{~j} 
\end{equation}
and compute 
\begin{subequations} \label{invariants}
\begin{equation} \label{I}
\I \equiv \frac{1}{2} \, C^i_{~j} C^j_{~i}
\end{equation}
and
\begin{equation} \label{J}
\J \equiv \frac{1}{6} \, C^i_{~j} C^j_{~k} C^k_{~i}.
\end{equation}
\end{subequations}
The four invariants of the Weyl tensor are then given by the real and imaginary parts of $\I$ and $\J$.  We note that $\I$ has units of $\lambda^{-4}$, while $\J$ has units of $\lambda^{-6}$.  For algebraically special spacetimes the invariants are related by
\begin{equation} \label{special}
    \I^3 = 27 \J^2
\end{equation}
(see \cite{SteKMHH03}).   We also note that the Kretschmann scalar 
\begin{equation}
\K \equiv {}^{(4)} R^{abcd} \, {}^{(4)} R_{abcd}
\end{equation}
is related to the real part of $\I$ by $\K = 16 \Re(\I)$.

For our axisymmetric and twist-free spacetimes, the only non-zero components of $\B_{ij}$ are the $(r\varphi)$ and $(\theta\varphi)$ components, while, for $\E_{ij}$, these components vanish identically.  As a result, both $\I$ and $\J$ are real.  We furthermore observe that the relation (\ref{special}) holds for our spacetimes on the symmetry axis, but not, in general, elsewhere.  As an example we show the two curvature scalars for our initial data in Fig.~\ref{fig:indata}.

In addition to evaluating the curvature scalars locally in space and time, it is also useful to consider measures of the curvature that depend on time only.  One option are the values of $\I$ and $\J$ at the center, which, as we discussed above, represents a preferred observer.  However, the innermost grid points (i.e.~those for the smallest value of the radius) in our code using spherical polar coordinates are the ones that are most strongly affected by numerical error, which leads to noticeable artifacts in $\I$ and $\J$ at late times.  We therefore consider two different alternatives that, as an added benefit, provide global information about the curvature invariants.   Specifically, we compute the local-in-time maximum magnitudes of the invariants, i.e.
\begin{equation} \label{maxima}
\I_{\rm max}(t) \equiv \max_{\Sigma} | \I(t,r,\theta) |,
\end{equation}
and similarly for $\J$.  Here the maximum is taken over the current spatial slice $\Sigma$ of constant coordinate time $t$, except that we disregard the grid points for the smallest values of the radius $r$ in order to reduce the effects of numerical error near the origin.  At late times, however, when numerical error affects more than just the innermost grid points, this measure becomes unreliable as well.  As an alternative we also consider the spatial proper integrals over all space
\begin{equation} \label{integrals}
    \I_{\rm int}(t) \equiv \int_\Sigma | \I | dV,
\end{equation}
and similar for $\J$. Unlike for the maxima in (\ref{maxima}), we {\em do} include the innermost grid points in the integrals (\ref{integrals}).  We also note that $\I_{\rm int}$ has units of $\lambda^{-1}$, while $\J_{\rm int}$ has units of $\lambda^{-3}$.

\section{Results}
\label{sec:results}

\subsection{Quadrupolar waves: $\ell = 2$}
\label{sec:quadru}

We start our discussion with quadrupolar ($\ell = 2$) waves, constructed from the expressions in Appendix \ref{sec:appA_l=2}, and adopt a number of different wave amplitudes $\A$ in the seed function (\ref{seed}).   All simulations shown in this section were performed with $N_r = 384$ (non-uniform) radial gridpoints that, initially, extend to the outer boundary at $r_{\rm out}^{\rm init} = 64$.  We allow regridding during these evolutions with a maximum ten-fold increase of resolution, in which case $r_{\rm out}^{\rm final} = 6.4$.  We also use $N_\theta = 96$ uniform, azimuthal angular grid points, covering one hemisphere.

\subsubsection{Fine-tuning to the threshold solution}
\label{sec:quadru:finetuning}

\begin{figure}[t]
    \centering
    \includegraphics[width = 0.45 \textwidth]{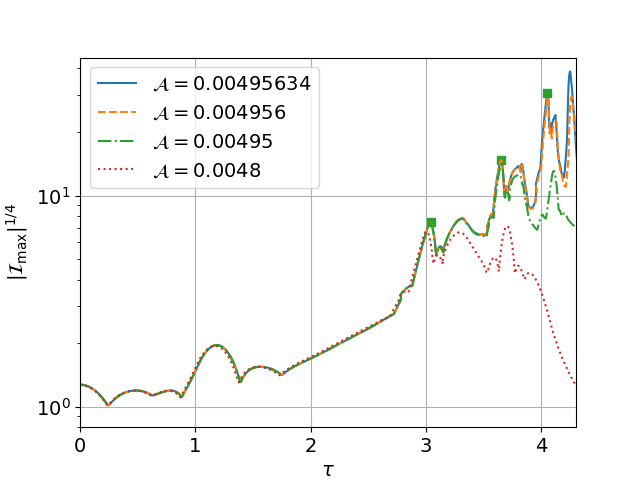}

    \includegraphics[width = 0.45 \textwidth]{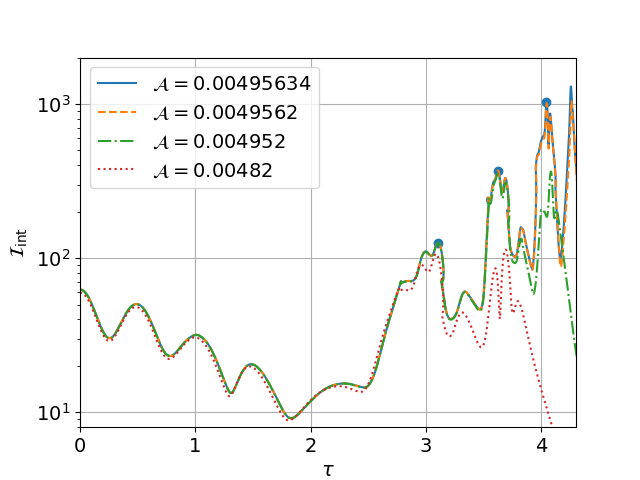}
    
     \caption{The maxima (\ref{maxima}) (top panel) and integrals (\ref{integrals}) (bottom panel) of $\I$ as a function of proper time $\tau$ (as measured at the center) for $\ell = 2$ waves with different amplitudes $\A$. In addition to our highest-amplitude subcritical data (for $\A = 0.00495634$) we selected amplitudes $\A$ for which a newly emerging peak in the curvature is approximately as high as the previous peak.  The dots and squares mark the times and values of the first three curvature peaks.}
    \label{fig:peeling_l_2}
\end{figure}

For sufficiently small amplitudes, the wave is subcritical, meaning that it disperses and leaves behind flat space, while, for sufficiently large amplitudes, it is supercritical, meaning that it collapses to form a black hole.\footnote{While we refer to the limiting solution as the {\em threshold} solution, rather than the {\em critical} solution, in order to emphasise that it is not universal, we continue to use the terms supercritical, subcritical, and near-critical to describe whether or not a black hole is formed.}  For the former, the maximum values of the curvature invariants $\I$ and $\J$ decay after attaining some extremum over the course of the evolution, while, for the latter, they diverge and approach infinity.   We also observe that, at late times, the minimum value of the lapse function $\alpha$ approaches unity for subcritical waves, but drops to small values and then performs oscillations around zero for supercritical waves (see \cite{LiBDdeO23} for a discussion of the origins of the oscillations).  A similar ``collapse of the lapse" has been observed for a number of other slicing conditions (see, e.g., \cite{BeiO98} for an analytical study of maximal slicing), but a priori it is not clear that the shock-avoiding slicing condition employed here would display a similar behavior.  

As an example we show in Fig.~\ref{fig:peeling_l_2} the maxima (\ref{maxima}) and integrals (\ref{integrals}) of the curvature scalar $\I$ as a function of proper time $\tau$ for selected subcritical values of the amplitude $\A$.  Note that an increasing number of peaks emerges at late time as $\A$ is fine-tuned to its critical value.  Fine-tuning the amplitude $\A$ to the onset of black-hole formation we bracket the critical amplitude $\A_*$ to approximately $0.00495634 < \A_* < 0.00495637$.  While our computational methods and resources limit us to a more modest fine-tuning than some other recent studies of critical collapse of gravitational waves (see \cite{LedK21,FerRABH22}), it did allow us to observe several properties of critical Teukolsky waves, as we will discuss in the following.  

\subsubsection{The threshold solution: maxima}
\label{sec:quadru:properties}

We start by observing that, after about $\tau \simeq 2.5$, the graphs in Fig.~\ref{fig:peeling_l_2} show patterns that repeat with increasing frequency and increasing amplitude, as is expected for a DSS solution.  We can follow these repeated patterns for about two or three periods, after which neither our fine-tuning nor our grid resolution is sufficient to reliably model the threshold solution.

In order to analyze the properties of the threshold solution quantitatively we introduce two new {\em similarity coordinates} adapted to self-similarity, namely a new time coordinate
\begin{equation} \label{slowtime}
    T \equiv - \ln(\tau_* - \tau),
\end{equation}
which we will refer to as ``slow time", and a dimensionless rescaled radial coordinate
\begin{equation} \label{xi}
    \xi \equiv R / (\tau_* - \tau),
\end{equation}
where $R$ is the proper distance from the center measured along the time slice of constant $t$.  In both (\ref{slowtime}) and (\ref{xi}) $\tau_*$ denotes the proper time of the accumulation event as measured by the observer at the center.  Note also that (\ref{xi}) assumes that the accumulation event is located at the center.  A dimensionless quantity describing a continuously self-similar contracting solution is a function of $\xi$ only;   this means that a spatial feature will appear at distance $R$ from the origin that is proportional to $\tau_* - \tau$.  Similarly, a quantity with dimension $\lambda^n$ will scale with $R^n \simeq (\tau_* - \tau)^n$. 
For some matter models the self-similarity is discrete rather than continuous.  An exact DSS has been observed, for example, in the spherically symmetric collapse of scalar fields \cite{Cho93,Gun97}, while an approximate DSS has been reported for the gravitational collapse of electromagnetic wave \cite{BauGH19,PerB21}.  Here we argue that such an approximate DSS exists for the collapse of gravitational waves as well.  The periodicity of a DSS solution is described by the {\em echoing period} $\Delta$ in the slow time $T$.

The accumulation time $\tau_*$ and the period $\Delta$ can be determined in a number of different ways that, for an exact DSS, for perfect fine-tuning, and in the absence of numerical error would all result in identical values.  First note that the period $\Delta$ can be written as
\begin{equation}\label{Delta1}
    \Delta = - \ln \frac{\tau_* - \tau_{i+1}}{\tau_* - \tau_i}
    = - \ln \frac{\tau_* - \tau_{i}}{\tau_* - \tau_{i-1}},
\end{equation}
where $\tau_{i+1}$, $\tau_i$,and $\tau_{i-1}$ are the proper times of three subsequent maxima. These two equations can be solved for
\begin{equation} \label{tau_star}
    \tau_* = \frac{\tau_{i-1} \tau_{i+1} - \tau_{i}^2}{\tau_{i-1} - 2 \tau_{i} + \tau_{i+1}}, \\
\end{equation}
and
\begin{equation} \label{Delta1bis}
    \Delta = \ln\frac{\tau_i-\tau_{i-1}}{\tau_{i+1}-\tau_i}
\end{equation}
in terms of the three observed maxima.

Alternatively, we can consider a quantity, say ${\mathcal Q}$, with dimension $\lambda^n$.  Since such a quantity should be proportional to $(\tau_* - \tau)^{n}$, the ratio of its values at two subsequent maxima should satisfy
\begin{equation} 
    \frac{{\mathcal Q}_{i+1}}{{\mathcal Q}_i} = \left( \frac{\tau_* - \tau_{i+1}}{\tau_* - \tau_{i}} \right)^n =
    e^{-n \Delta}
\end{equation}
where we have used (\ref{Delta1bis}) in the last step, or
\begin{equation} \label{Delta2}
    \Delta = - \frac{1}{n} \ln \left( {\mathcal Q}_{i+1} / {\mathcal Q}_i \right). 
\end{equation}
We see that (\ref{Delta2}) provides an estimate of $\Delta$ that is based on the values of the maxima alone.  Inserting this value of $\Delta$ back into the first equality in (\ref{Delta1}), together with the corresponding times $\tau_i$ and $\tau_{i+1}$, then yields 
\begin{equation} \label{tau_star2}
    \tau_* = \frac{{\mathcal Q}_{i+1}^{1/n} \tau_i - {\mathcal Q}_{i}^{1/n} \tau_{i+1}}{{\mathcal Q}_{i+1}^{1/n} - {\mathcal Q}_{i}^{1/n}}
\end{equation}
as another estimate for $\tau_*$.

Adopting the values marked by the dots in the top panel of Fig.~\ref{fig:peeling_l_2}, i.e.~for the curvature scalar $\I_{\rm max}$, we find $\tau_* \simeq 4.8$ and $\Delta \simeq 0.42$ from the times $\tau_i$ of the maxima, using (\ref{tau_star}) and (\ref{Delta1bis}).  Using (\ref{Delta2}) for the values of the maxima, on the other hand, we find $\Delta \simeq 0.68$ and $0.74$ for the first and second pair, corresponding to $\tau_* \simeq 4.3$ and $4.0$
from (\ref{tau_star2}).  We obtain similar values for the maxima of $\J_{\rm max}$ as well as $\J_{\rm int}$.  There are several possible reasons for the two approaches yielding somewhat different values, including our modest fine-tuning.  More importantly, however, we would expect to obtain the exact same values only for an {\em exact} DSS, which we do not believe is the case here.  

Using the maxima of $\I_{\rm int}$ (marked by the dots in the lower panel of Fig.~\ref{fig:peeling_l_2}), the differences in the values of $\Delta$ and $\tau_*$ obtained from the two approaches differ even more.  Using (\ref{Delta1}) we find $\tau_* = 5.8$ and $\Delta = 0.21$, while from (\ref{Delta2}) we obtain $\Delta \simeq 1.06$ and $1.05$ for the two pairs, corresponding to $\tau_* \simeq 3.9$ and 4.4. We take these inconsistencies as indication that, at least in the regime considered here, $\I_{\rm int}$ is less a reliable measure than the maximum values of the curvature.  One possible reason for this is that $\I_{\rm int}$ is not, or not yet, dominated by the threshold solution inside the past light-cone of the accumulation event.  Furthermore, for $\ell = 2$ the maxima of $\I_{\rm max}$ occur at the center, so that its values and times have a gauge-independent meaning, while the values of $\I_{\rm int}$ depend on the slicing.

\begin{figure}[t]
    \centering
    \includegraphics[width = 0.45 \textwidth]{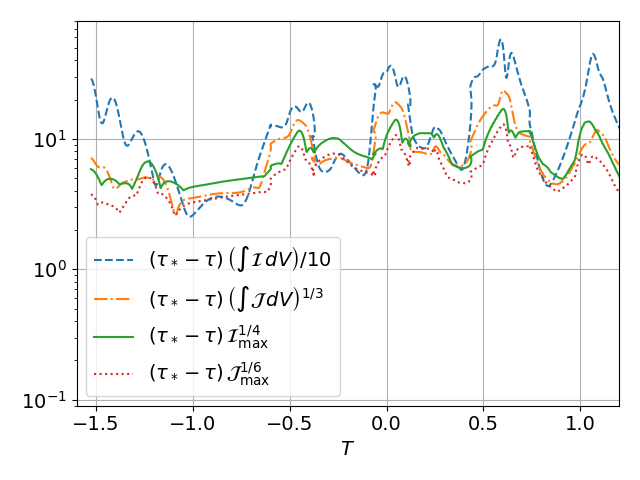}
    \caption{Curvature measures as a function of slow time $T = -\log(\tau_* - \tau)$ for the near-critical $\ell = 2$ wave with amplitude $\A = 0.00495634$ and assuming $\tau_* = 4.6$.}
    \label{fig:crit_sol_l_2}
\end{figure}

As a compromise we now adopt $\tau_* = 4.6$ in the following analysis.  In particular, we show in Fig.~\ref{fig:crit_sol_l_2} the maxima (\ref{maxima}) and integrals (\ref{integrals}) of the curvature invariants $\I$ and $\J$ as a function of slow time $T$.  We take an appropriate root of each measure to obtain a quantity with units of $\lambda^{-1}$, and then multiply with $\tau_* - \tau$ so that, for exact self-similarity, the resulting curve would be exactly periodic.  As in the previous figures the plot is dominated by the initial data at early times, and by a decrease in the curvature at late times, when the (sub-critical) wave disperses.  At intermediate times, however, all our curvature measures display an approximate periodicity with a period of about $\Delta \simeq 0.5$, as expected from our discussion above, consistent with an approximate self-similarity of the threshold solution.  Interestingly, this value of $\Delta$ is consistent with those reported by \AE\ (see Table I in \cite{AbrE94}), who also considered Teukolsky-wave initial data, albeit not time-symmetric ones. 

\subsubsection{The threshold solution: spacetime diagrams}
\label{sec:quadru:spacetime}

\begin{figure}[t]
    \centering
    \includegraphics[width = 0.45 \textwidth]{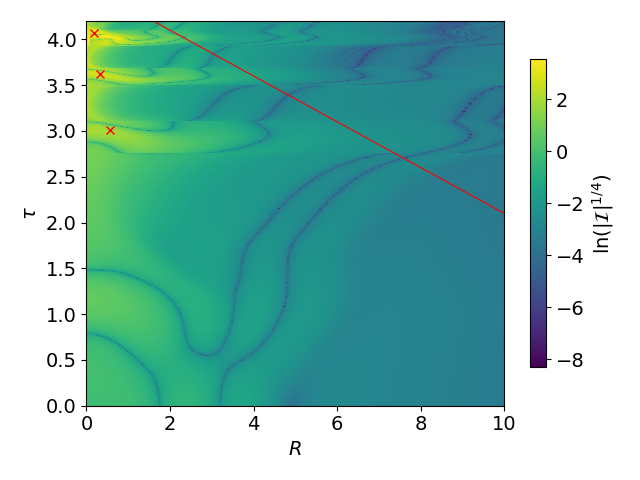}
    
    \includegraphics[width = 0.45 \textwidth]{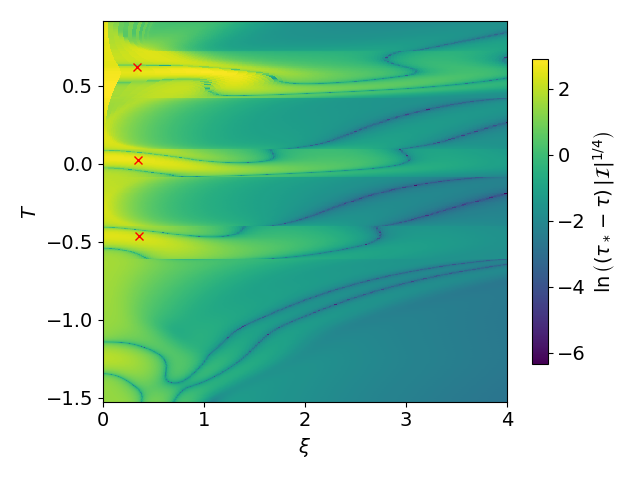}

    \caption{The Weyl curvature scalar $\I$ for a near-critical $\ell = 2$ wave with amplitude $\A = 0.00495634$ on the symmetry axis (corresponding to the $z$-axis in Fig.~\ref{fig:indata}).  The top panel shows $\I^{1/4}$ as a function of proper distance $R$ from the origin and of proper time $\tau$ as measured by an observer at the origin.  In the bottom panel we show $(\tau_* - \tau) \I^{1/4}$ as a function of the similarity coordinates $\xi$ and $T$ assuming $\tau_* = 4.6$ (see \eqs~\ref{slowtime} and \ref{xi}).  We include the red crosses as guidance in identifying corresponding features in the two plots.  The red line in the top panel marks the outer boundary at $\xi = 4$ in the lower panel.  (The left, lower, and upper boundaries of the two plots correspond to each other.)}
    \label{fig:crit_sol_axis_l_2}
\end{figure}

\begin{figure*}
    \centering
    \includegraphics[width = 0.45 \textwidth]{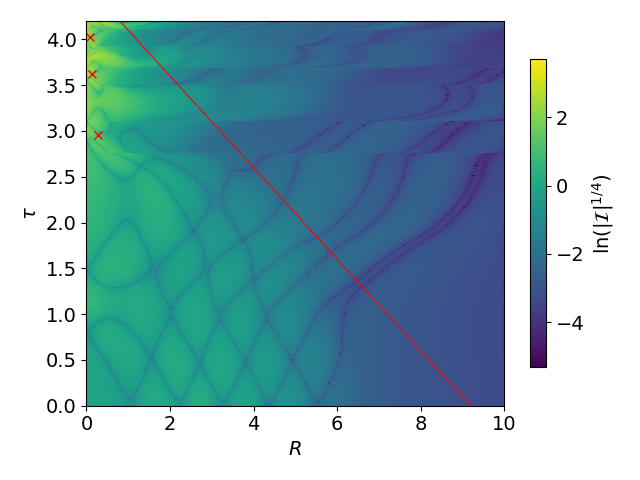}
    \includegraphics[width = 0.45 \textwidth]{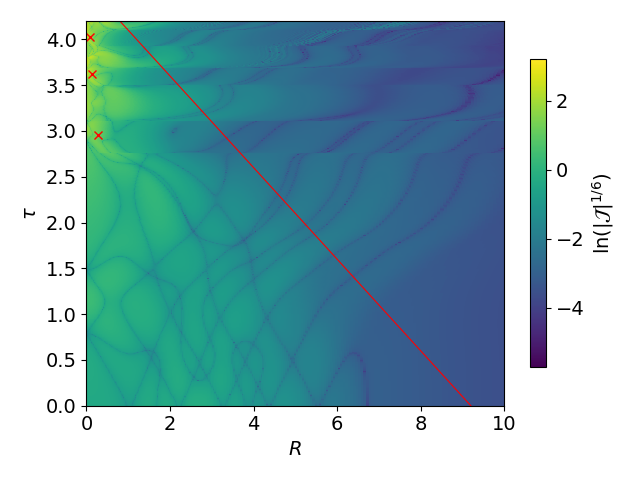}

    \includegraphics[width = 0.45 \textwidth]{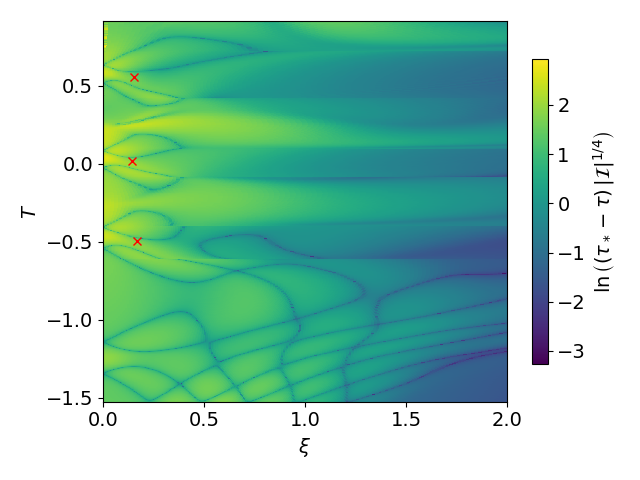}
    \includegraphics[width = 0.45 \textwidth]{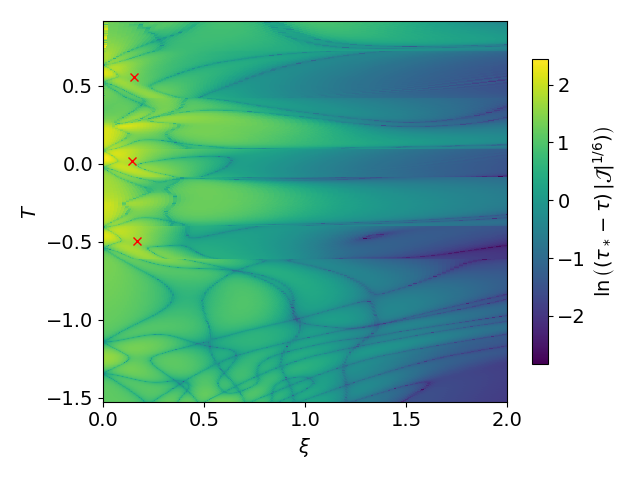}

    \caption{The Weyl curvature scalars $\I$ (left column) and $\J$ (right column) for a near-critical $\ell = 2$ wave with amplitude $\A = 0.00495634$ in the equatorial plane (corresponding to the $x$-axis in Fig.~\ref{fig:indata}). As in Fig.~\ref{fig:crit_sol_axis_l_2} we show the data as functions of $R$ and $\tau$ in the top row, and as functions of the similarity coordinates $\xi$ and $T$ in the bottom row, we include red crosses to identify similar patterns, and the red lines in the top panels mark the outer boundaries of the bottom panels (see text for details and discussion).}
    \label{fig:crit_sol_eq_l_2}
\end{figure*}

It is also useful to study the behavior of the threshold solution in the entire spacetime, rather than only its maxima or integrals.  We therefore show in Figs.~\ref{fig:crit_sol_axis_l_2} and \ref{fig:crit_sol_eq_l_2} the curvature invariants for $\A = 0.00495634$, i.e.~for our highest-amplitude subcritical wave.  

In Fig.~\ref{fig:crit_sol_axis_l_2} we plot $\I$ on the symmetry axis, i.e.~in the direction of the $z$-axis as shown in Fig.~\ref{fig:indata}.  We only show $\I$ in this figure, because our data satisfy (\ref{special}) on the axis, so that $\I$ and $\J$ are not independent there.  We display the data in two different ways, however.  In the top panel we show (the logarithm of) $\I^{1/4}$ (with units of $\lambda^{-1}$) as a function of proper radius $R$ from the origin (along a slice of constant coordinate time), and of proper time $\tau$ as measured at the center.  As expected from our previous discussion we observe, after $\tau \simeq 2.5$, features that appear to repeat with increasing frequency and on shorter length scales, again suggesting an approximate DSS with an accumulation event at the center.  We include the red crosses in the figures to help the reader identify these features; we will discuss the choice of the location of these crosses in more detail at the end of this subsection (they are placed at values of $T_\text{null}$ spaced by $\Delta=0.53$ and constant values of $\lambda$).

In the bottom panel of Fig.~\ref{fig:crit_sol_axis_l_2} we show (the logarithm of) the dimensionless combination $(\tau_* - \tau) \, \I^{1/4}$ as a function of the similarity coordinates $\xi$ and $T$.  We observe that the solution indeed features similar patterns that repeat in slow time with a periodicity $\Delta \simeq 0.53$. 

We also note an artifact in the figures that appears to suggest that the curvature invariants change discontinuously at certain specific times, for example around $\tau = 2.7$.  This effect is a consequence of the lapse function dipping below zero at the center at certain times (see Fig.~\ref{fig:slicing}), so that the proper time advances {\em backwards} at the center, while, at sufficient distance away from the center where the lapse is still positive, proper time still advances forward.  Displaying the data as a function of proper time as measured by an observer at the center therefore leads to discontinuities away from the center.  Another artifact appears in the bottom panel of 
Fig.~\ref{fig:crit_sol_axis_l_2} and similar figures below for large values of $T$ and small values of $\xi$; these artifacts result from insufficient resolution for the increasingly fine structures that form at late times for near-critical solutions.

As it turns out, for quadrupolar waves the curvature invariants show significantly more structure in the equatorial plane than along the axis, as shown in Fig.~\ref{fig:crit_sol_eq_l_2}.  Since, away from the axis, $\I$ and $\J$ do not generally satisfy (\ref{special}), we now show $\I$ in the left column and $\J$ in the right column.  As in Fig.~\ref{fig:crit_sol_axis_l_2} we also show the scalars in terms of $R$ and $\tau$ in the top row, and in terms of the similarity coordinates $\xi$ and $T$ in the bottom row. 

At early times $\tau \lesssim 2$, and when plotted in terms of $R$ and $\tau$, the curvature scalar $\I$ displays a diagonal chess-board pattern, which is a consequence of the initial data representing a superposition of ingoing and outgoing waves.  At later times, $\tau \gtrsim 2.5$, we again observe the emergence of repeated patterns that occur with increasing frequency and on smaller length scales, before the wave disperses at late times.  The scalar $\J$ shows a similar but slightly more complicated behavior, which is perhaps not surprising given that even for the initial data, as shown in Fig.~\ref{fig:indata}, it appears somewhat more complicated.  

Plotting $\I$ and $\J$ in terms of the similarity coordinates $\xi$ and $T$, as shown in the bottom row of Fig.~\ref{fig:crit_sol_eq_l_2}, reveals that that repeating patterns at late times are again nearly periodic in $T$ with a period of about $\Delta \simeq 0.53$, suggesting the existence of an at least approximate discrete self-similarity with an accumulation event at the center.  

\begin{figure}[t]
    \centering
    \includegraphics[width = 0.45 \textwidth]{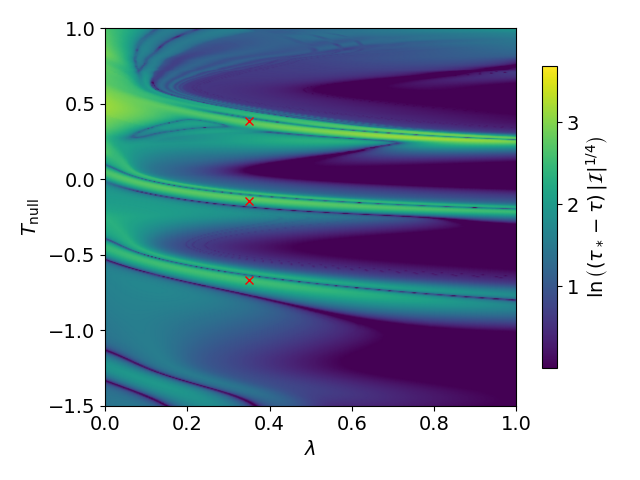}
    \caption{The curvature scalar $\I$ for a near-critical $\ell = 2$ wave along the symmetry axis, as in Fig.~\ref{fig:crit_sol_axis_l_2}, but with $\I$ as observed along outgoing null geodesics (see text for details).}
    \label{fig:photon_plot_axis_l_2}
\end{figure}

\begin{figure}[t]
    \centering
    \includegraphics[width = 0.45 \textwidth]{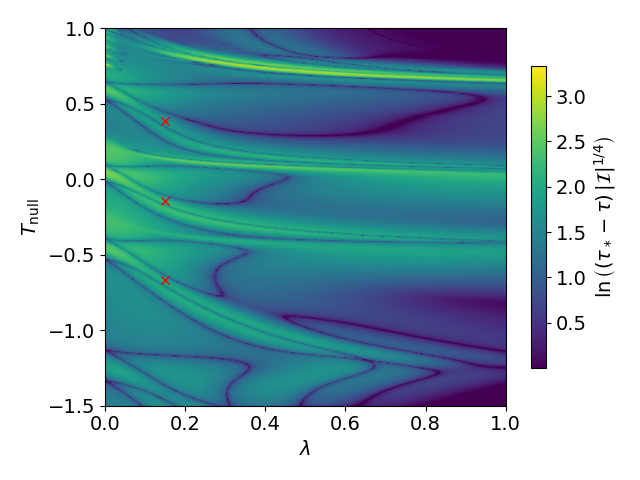}
    
    \includegraphics[width = 0.45 \textwidth]{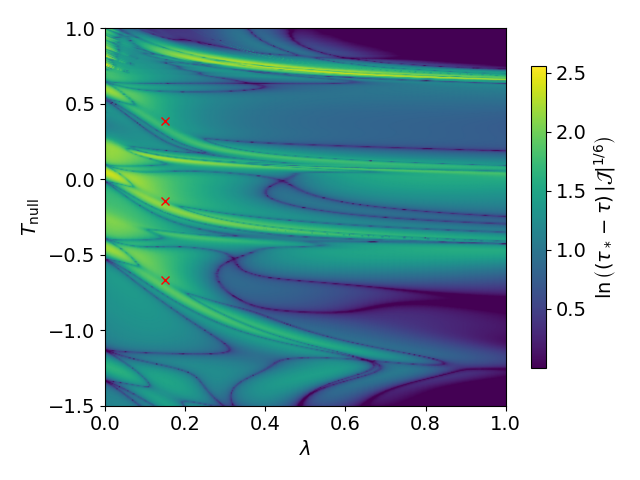}
    \caption{The curvature scalars $\I$ (top panel) and $\J$ (bottom panel) for a near-critical $\ell = 2$ wave in the equatorial plane, as in Fig.~\ref{fig:crit_sol_eq_l_2}, but with $\I$ and $\J$ as observed along outgoing null geodesics.}
    \label{fig:photon_plot_eq_l_2}
\end{figure}

In twist-free axisymmetry with an additional equatorial symmetry, within any time slice compatible with that symmetry, the world sheets of the $x$ and $z$ axes are already geometrically unique. However, there still is a gauge-dependence of the plots shown in Figs.~\ref{fig:crit_sol_axis_l_2} and \ref{fig:crit_sol_eq_l_2} through  the time slicing: the value of $\tau$ assigned to an event is the value on the slice through it, and its proper distance $R$ is integrated along the slice.  The fact that these figures display any periodic behavior away from the center at all suggests that the shock-avoiding slicing condition leads to coordinates that reflect the self-similarity to some degree. 

As an alternative, we can also construct null coordinates on the $x$ and $z$-axis world sheets that have a completely gauge-invariant meaning.  Specifically, we consider null geodesics emitted from the center, both in the axial and the equatorial direction.  We label the null geodesics by the slow time $T_{\rm null}$ at which they are emitted from the center, and an affine parameter $\lambda$ normalized by $\lambda = 0$ and $(d\lambda/d\tau)_{T_{\rm null}} = dT/d\tau = (\tau_* - \tau)^{-1}$  at the center (so that, initially, $\lambda$ advances at the same rate as $T$).  We show plots of the curvature invariants on the axis and on the equator in Figs.~\ref{fig:photon_plot_axis_l_2} and \ref{fig:photon_plot_eq_l_2}.  Again, these plots show approximately periodic features at intermediate times, i.e.~after the initial data have evolved toward a threshold solution, and before the latter disperses to infinity.  

In order to highlight this self-similarity we again include red crosses in these figures.  Specifically, we pick $\lambda = 0.35$ for the plots on the axis and $\lambda = 0.15$ for those on the equator, and choose values of $T_{\rm null} = -0.67$, $-0.14$, and $0.39$, which differ by multiples of $\Delta = 0.53$ and correspond to similar phases in the oscillations.  The crosses in Figs.~\ref{fig:crit_sol_axis_l_2} and \ref{fig:crit_sol_eq_l_2} correspond to the same spacetime events as those in Figs.~\ref{fig:photon_plot_axis_l_2} and \ref{fig:photon_plot_eq_l_2}.  For an exact DSS, for the exact values of $\tau_*$ and $\Delta$, $\log(\tau_* - \tau)\I^{1/4}$ would be exactly periodic in $T_\text{null}$ at constant $\lambda$, and similarly for $\J$.

If our time slicing were also compatible with an exact DSS, the same would be true for our earlier similarity coordinates $T$ and $\xi$.  In particular, the crosses would appear at the same values of $\xi$ in the lower panels of Figs.~\ref{fig:crit_sol_axis_l_2} and \ref{fig:crit_sol_eq_l_2}.  The fact that they do indeed appear at similar locations suggests that all three of the above conditions are met approximately.

\subsubsection{Scaling with distance from the black hole threshold}
\label{sec:quadru:scaling}
 
\begin{figure}[t]
    \centering
    \includegraphics[width = 0.45 \textwidth]{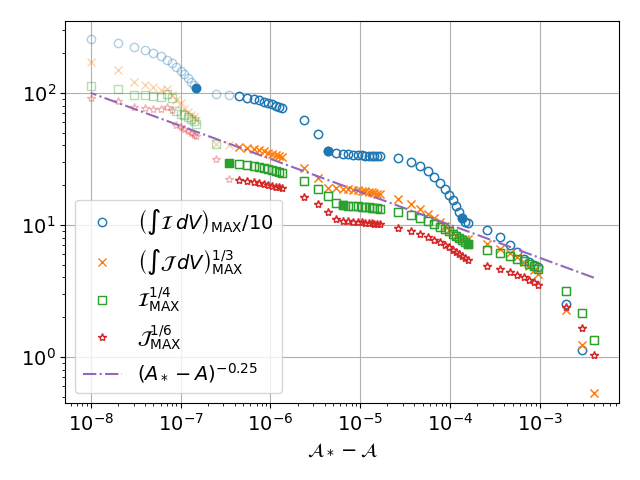}
    \caption{Global measures of the maximum curvature attained for quadrupolar $\ell = 2$ waves as a function of amplitude ${\mathcal A}$, assuming $\A_* = 0.00495635$.  The filled symbols correspond to the amplitudes shown in Fig.~\ref{fig:peeling_l_2} for which a new peak in the curvature emerges and starts dominating over a previous peak; these amplitudes therefore correspond to a ``kink" in the scaling plot shown here.  Results for near-critical amplitudes, for which we believe that numerical error close to the origin leads to an overestimate in our values of the curvature maxima, are included as shaded symbols only.}
    \label{fig:scaling_l_2}
\end{figure}

We now change focus, from our most fine-tuned subcritical solution (as an approximation to the threshold solution) to the scaling of the curvature measures for subcritical solutions as a function of distance $\A_*- \A$ from the black hole threshold.  Specifically, we now compute the (global) maxima of the curvature measures (\ref{maxima}) and (\ref{integrals}) over time, e.g.
\begin{equation} \label{maxmax}
    \I_{\rm MAX} \equiv \max_t \, \I_{\rm max}(t),
\end{equation}
and similar for $\J$ as well as their integrals (\ref{integrals}).  We use capital letters in the subscript in order to distinguish the global maxima from the local-in-time maxima, denoted with lower-case subscripts.  In Fig.~\ref{fig:scaling_l_2} we show these global curvature maxima, properly rescaled so that all have units of $\lambda^{-1}$, as a function of the wave amplitude $\A$, assuming $\A_* = 0.00495635$.  Because of our limited numerical resolution we found that several of our curvature measures become quite noisy at late times when $\A$ is too close to its critical value, and we therefore shaded the corresponding data in Fig.~\ref{fig:scaling_l_2}.  We also marked the simulations for the amplitudes shown in Fig.~\ref{fig:peeling_l_2} as filled symbols.  Since, for these amplitudes, a newly emerging curvature peak starts to dominate over previous peaks, these amplitudes correspond to the ``kinks" in the scaling plot of Fig.~\ref{fig:scaling_l_2}.

For a unique critical solution with {\em continuous} self-similarity and a single unstable mode one would expect a curvature measure with units of $\lambda^{-1}$, say ${\mathcal C}$, to scale with
\begin{equation} \label{scaling}
    {\mathcal C} \simeq |\A_* - \A|^{- \gamma},
\end{equation}
where $\gamma$ is the inverse of the mode's Lyapunov exponent (see, e.g., \cite{KoiHA95,Mai96,GarD98}).  For a DSS critical solution one expects a periodic ``wiggle" with period of
\begin{equation} \label{wiggleperiod}
    {\mathcal P} \equiv \left| \ln \left( \frac{\A_* - \A_{i+1}}{\A_* - \A_i} \right) \right| = \frac{\Delta}{\gamma}
\end{equation}
in $\ln|\A_*-A|$ to be superimposed on the power-law scaling (\ref{scaling}) (see \cite{Gun97,HodP97}).  

If we were to estimate $\gamma$ as the slope of the straight line connecting two neighbouring kinks (the full blue dots or full green squares) in Fig.~\ref{fig:scaling_l_2} and substitute this value into Eq.~(\ref{wiggleperiod}), we would by obtain a value of $\Delta$ very similar to (\ref{Delta2}) -- the only difference being that, in Fig.~\ref{fig:peeling_l_2}, we identified the maxima from the threshold solution, while, in Fig.~\ref{fig:scaling_l_2}, the kinks are the maxima of subcritical solutions. However, if we estimate $\gamma$ independently as a best fit line through the observed wiggly scaling laws, \eq~(\ref{wiggleperiod}) provides other estimates of $\Delta$.

In the absence of an exact DSS, and given our modest fine-tuning, it is difficult to determine the critical exponent $\gamma$ precisely.  Focusing on the curvature maxima, which we have previously identified as being a more reliable diagnostics than the integral of $\I$, we see that the data are reasonably well fitted by $\gamma = 0.25$.  This value is slightly smaller than those found by \cite{LedK21}, who estimate $\gamma = 0.33$ for their (positive-amplitude) Teukolsky data, and \AE, who report $\gamma \simeq 0.36$, but the initial data of both \cite{LedK21} and \AE\ are also different from ours in that they are not time-symmetric.

Measuring the distance between the ``kinks" in the data for $\I_{\rm MAX}$ in Fig.~\ref{fig:scaling_l_2} we estimate ${\mathcal P} \simeq 3$.  Using $\gamma \simeq 0.25$, we then obtain $\Delta \simeq 0.75$ from (\ref{wiggleperiod}), which is consistent with our earlier estimate from the maximum values of $\I$ in (\ref{Delta2}).

\subsection{Hexadecapolar waves: $\ell = 4$}
\label{sec:hexadeca}

\begin{figure}[t]
    \centering
    \includegraphics[width = 0.45 \textwidth]{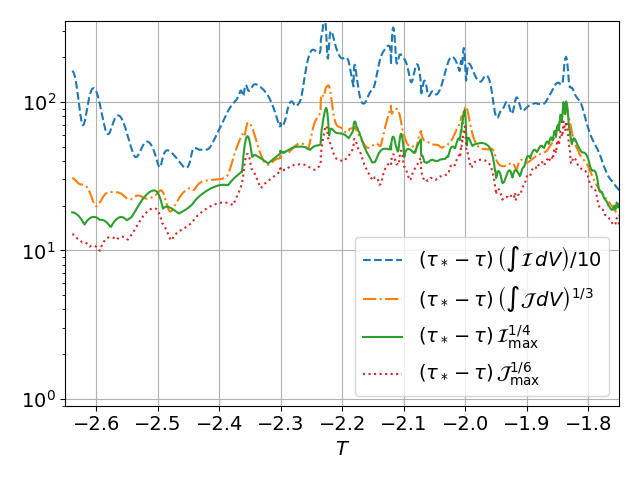}
    \caption{Curvature measures as a function of slow time $T = -\log(\tau_* - \tau)$ for the near-critical $\ell = 4$ wave with amplitude $\A = 4.251 \times 10^{-4}$ and assuming $\tau_* = 14$.}
    \label{fig:crit_sol_l_4}
\end{figure}

\begin{figure}[t]
    \centering
    \includegraphics[width = 0.45 \textwidth]{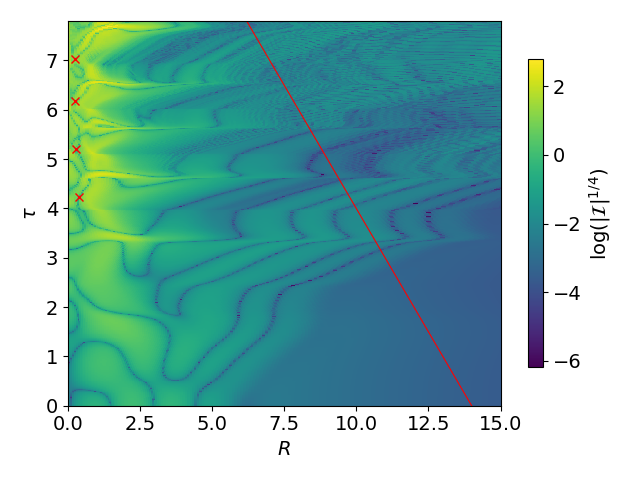}
    
    \includegraphics[width = 0.45 \textwidth]{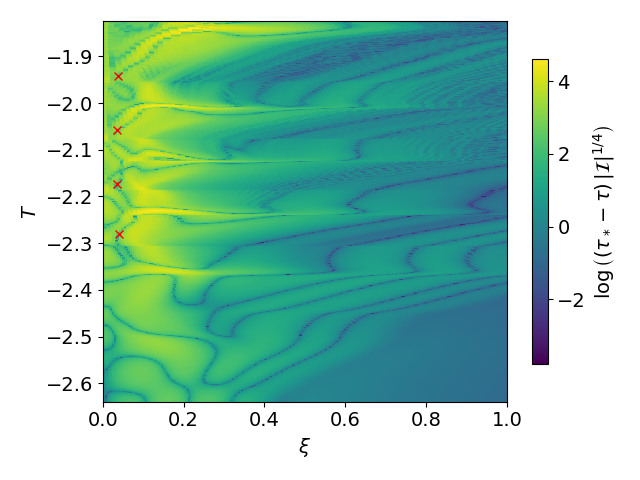}

    \caption{The Weyl curvature scalar $\I$ for a near-critical hexadecapolar $\ell = 4$ wave with amplitude $\A = 4.251 \times 10^{-4}$ on the symmetry axis (compare with Fig.~\ref{fig:crit_sol_axis_l_2} for quadrupolar waves).  The top panel shows $\I^{1/4}$ as a function of proper distance $R$ from the origin and of proper time $\tau$ as measured by an observer at the origin.  In the bottom panel we show $(\tau_* - \tau) \I^{1/4}$ as a function of the similarity coordinates $\xi$ and $T$ assuming $\tau_* = 14$ (see \eqs~\ref{slowtime} and \ref{xi}).  As in Fig.~\ref{fig:crit_sol_axis_l_2} and \ref{fig:crit_sol_eq_l_2} we include the red crosses as guidance in identifying corresponding features in the two plots, and the red line in the top panel marks the outer boundary at $\xi = 1$ in the lower panel.}
    \label{fig:crit_sol_axis_l_4}
\end{figure}

\begin{figure*}
    \centering
    \includegraphics[width = 0.45 \textwidth]{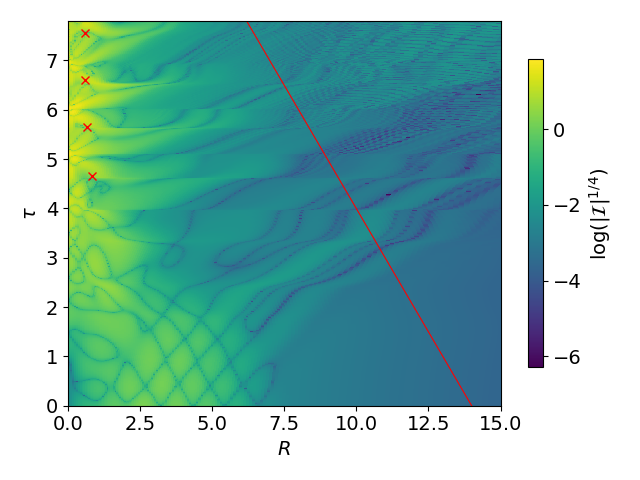}
    \includegraphics[width = 0.45 \textwidth]{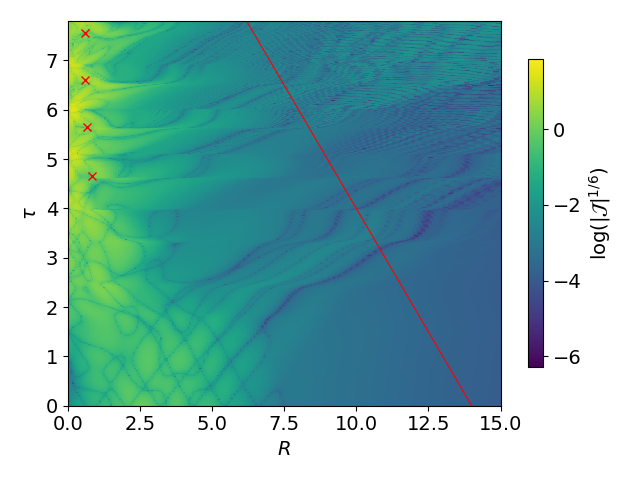}

    \includegraphics[width = 0.45 \textwidth]{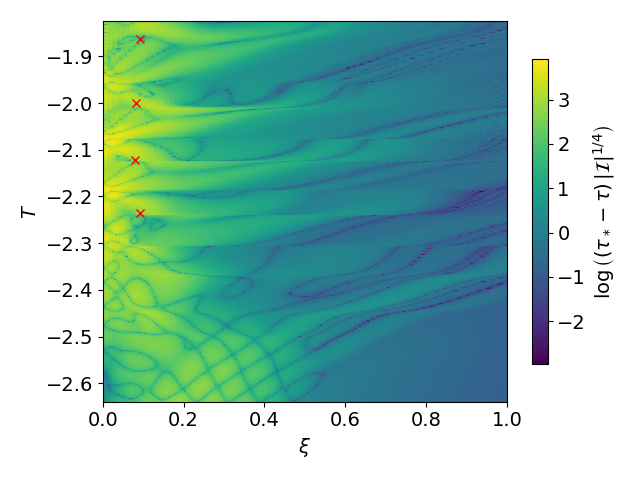}
    \includegraphics[width = 0.45 \textwidth]{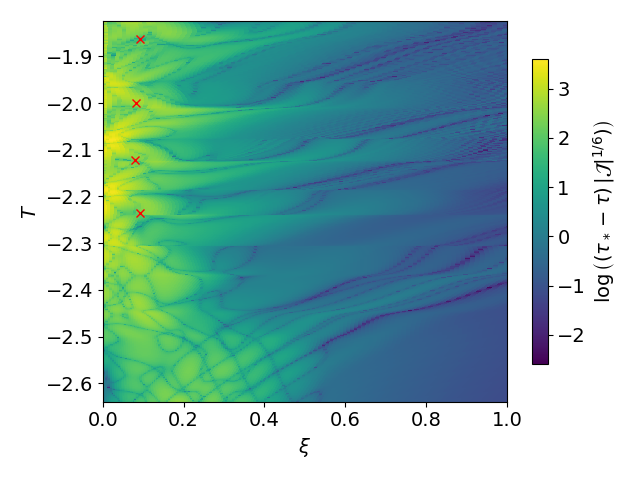}

    \caption{The Weyl curvature scalars $\I$ (left column) and $\J$ (right column) for a near-critical $\ell = 4$ wave with amplitude $\A = 4.251 \times 10^{-4}$ in the equatorial plane (compare with Fig.~\ref{fig:crit_sol_eq_l_2} for quadrupolar waves).   As in Fig.~\ref{fig:crit_sol_axis_l_4} we show the data as functions of $R$ and $\tau$ in the top row, and as functions of the similarity coordinates $\xi$ and $T$ in the bottom row (see text for details and discussion).}
    \label{fig:crit_sol_eq_l_4}
\end{figure*}

We now turn to hexadecapolar waves ($\ell = 4$), which we construct from the expressions in Appendix \ref{sec:appA_l=4}.  Given the more complicated angular structure of these data in comparison to the quadrupolar data (see Fig.~\ref{fig:indata}) we use $N_\theta = 128$ angular grid points for these simulations, but reduce the number of radial grid points to $N_r = 256$ in order to shorten the runtime of the simulations.  As we will discuss in more detail below, we also find that regridding is less helpful in these simulations than in those for quadrupolar data, and therefore keep the radial grid fixed during these evolutions. 

\subsubsection{Fine-tuning to the threshold solution, and its maxima}
\label{sec:hexadeca:finetuning}

As for the quadrupolar waves of Section \ref{sec:quadru}, we perform simulations for different amplitudes $\A$ in the seed function (\ref{seed}), and bracket the threshold amplitude to approximately $4.251 \times 10^{-4} < \A_* < 4.252 \times 10^{-4}$.   Corresponding to our (relatively) poorer resolution of the hexadecapolar waves, our fine-tuning is also worse than for the quadrupolar data.  We nevertheless include some results here in order to highlight some the qualitative differences between the threshold solutions for quadrupolar and hexadecapolar waves.

Fig.~\ref{fig:crit_sol_l_4} is the same as Fig.~\ref{fig:crit_sol_l_2}, except for hexadecapolar waves.  We again show curvature measures as a function of slow time $T$ for a near-critical solution.  Because our fine-tuning is even more modest than for the quadrupolar waves, it is even harder to determine the accumulation time $\tau_*$.  Here and in the following we will adopt $\tau_* = 14$ as a crude estimate.  Using this value of $\tau_*$ we see that, at intermediate times, the curvature measures show an approximate periodic behavior with a period of approximately $\Delta \simeq 0.1$, which is significantly shorter than that found for quadrupolar waves.  This behavior is similar to that of electromagnetic waves, for which \cite{PerB21} reported that the quadrupolar threshold solution has a shorter period than the dipolar threshold solution.

\subsubsection{The threshold solution}
\label{sec:hexadeca:spacetime}

In Figs.~\ref{fig:crit_sol_axis_l_4} and \ref{fig:crit_sol_eq_l_4}, which mirror Figs.~\ref{fig:crit_sol_axis_l_2} and \ref{fig:crit_sol_eq_l_2} for quadrupolar waves, we show the curvature scalars $\I$ and $\J$ on the axis and in the equatorial plane.  As an initial qualitative observation we note that, unlike the quadrupolar waves, which showed significantly more structure in the equatorial plane than on the axis, the hexadecapolar waves display rather complicated structure on the axis as well.  Also, while for quadrupolar waves the maximum values of the curvature scalars are typically found at the center, for the hexadecapolar waves they are typically away from the center, but on the symmetry axis. 

It is again intriguing to compare with results for the gravitational collapse of electromagnetic waves.  While \cite{BauGH19} found that the threshold solution for dipolar initial data features maximum density at the center, \cite{PerB21} reported that, for quadrupolar initial data, the threshold solution develops maximum densities on the symmetry axis, away from the center (see, e.g., Fig.~6 in \cite{PerB21}).   This qualitative difference between the threshold solutions for data with different multipoles is very similar to our findings for gravitational-wave data here.  

Unfortunately, the increasingly sharp spikes away from the center make it increasingly difficult to resolve them with the spherical polar coordinates used in our code.    While the regridding option in our code helps to zoom into center, is less effective for resolving features away from the center, as we discussed above.   We therefore do not attempt to examine the hexadecapolar waves more carefully here, and instead focus on qualitative differences from quadrupolar waves.

\begin{figure}[t]
    \centering
    \includegraphics[width = 0.45 \textwidth]{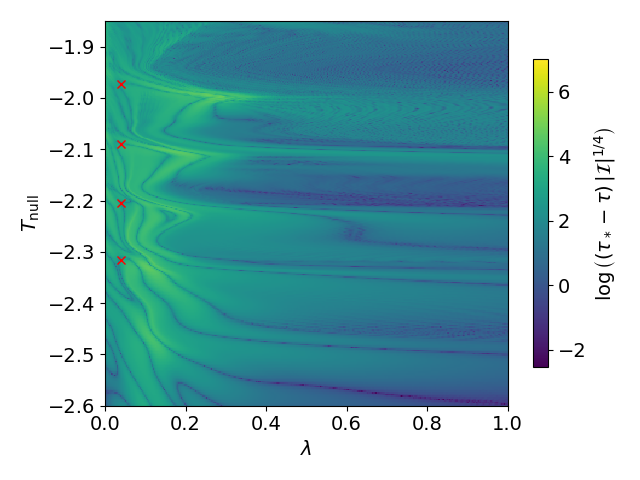}
    \caption{The curvature scalar $\I$ for a near-critical hexadecapolar $\ell = 4$ wave along the symmetry axis, as in Fig.~\ref{fig:crit_sol_axis_l_4}, but with $\I$ as observed along outgoing null geodesics (compare with Fig.~\ref{fig:photon_plot_axis_l_2} for quadrupolar data).}
    \label{fig:photon_plot_axis_l_4}
\end{figure}

\begin{figure}[t]
    \centering
    \includegraphics[width = 0.45 \textwidth]{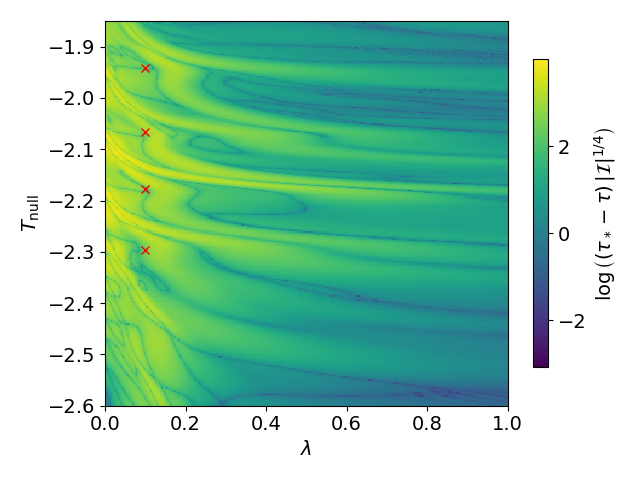}
    
    \includegraphics[width = 0.45 \textwidth]{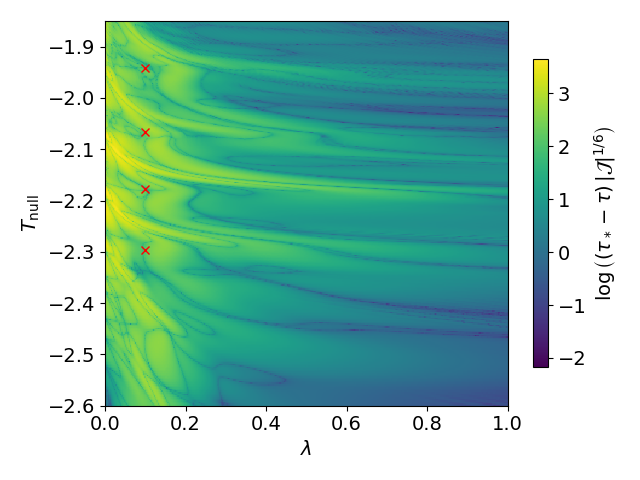}
    \caption{The curvature scalars $\I$ (top panel) and $\J$ (bottom panel) for a near-critical hexadecapolar $\ell = 4$ wave in the equatorial plane, as in Fig.~\ref{fig:crit_sol_eq_l_4}, but with $\I$ and $\J$ as observed along outgoing null geodesics (compare with Fig.~\ref{fig:photon_plot_eq_l_2} for quadrupolar data).}
    \label{fig:photon_plot_eq_l_4}
\end{figure}

In order to analyze whether the threshold solution for hexadecapolar waves features an approximate DSS (with an accumulation event at the center) we display the data as functions of the similarity coordinates (\ref{slowtime}) and (\ref{xi}) (see the bottom panels in Figs.~\ref{fig:crit_sol_axis_l_4} and \ref{fig:crit_sol_eq_l_4}), again with $\tau_* \simeq 14$.  While any DSS is certainly not exact, Figs.~\ref{fig:crit_sol_axis_l_4} and \ref{fig:crit_sol_eq_l_4} do suggest that there is an approximate DSS.  As before we include red crosses, constructed from the corresponding crosses in Figs.~\ref{fig:photon_plot_axis_l_4} and \ref{fig:photon_plot_eq_l_4} below, in order to help identify similar and repeated patterns.  Note that, despite the spikes forming away from the center, they appear to be features of an approximately DSS threshold solution with a single accumulation event at the center, rather than two separate centers of collapse -- similar to the threshold solution for quadrupolar electromagnetic waves discussed in \cite{PerB21}.  We again observe that the periodicity for these hexadecapolar waves is significantly shorter than that for the quadrupolar waves.  In fact, we were able to observe one more echo for the hexadocapolar waves than for the quadrupolar waves, despite the less accurate fine-tuning, precisely because the echoing period is shorter.  From Figs.~\ref{fig:crit_sol_axis_l_4} and \ref{fig:crit_sol_eq_l_4} see see that $\Delta \simeq 0.1$, as previously identified from Fig.~\ref{fig:crit_sol_l_4}, but we again caution that this value is affected by the estimate for $\tau_*$ and hence a crude estimate only.  Note that periodicity in $\tau$, rather than $T$, formally corresponds to the limit $\tau_*\to\infty$ and $\Delta\to 0$.  We can therefore estimate a small value of $\Delta$ less accurately than a large one.  Here, we cannot distinguish $\Delta\simeq 0.1$ from a periodicity in $\tau$ with certainty.  We are confident, however, that $\Delta\simeq 0.5$ is {\em not} a good fit to the $l=4$ threshold solution.

As for the quadrupolar waves, we also display the curvature scalars in terms of the gauge-invariant coordinates $T_{\rm null}$ and $\lambda$ constructed from outgoing null geodesics.  Specifically, we show the data along the symmetry axis in Fig.~\ref{fig:photon_plot_axis_l_4}, and in the equatorial plane in Fig.~\ref{fig:photon_plot_eq_l_4}.  While any DSS is certainly not exact, it is again easy to identify repeated patterns that suggest an approximate DSS.

Because of our poor fine-tuning for hexadecapolar waves, and because we have not sampled the subcritical regime with enough simulations to resolve kinks as in Fig.~\ref{fig:scaling_l_2}, we do not explore the scaling with distance from the black hole threshold in more detail, and do not attempt to estimate the critical exponent $\gamma$ for these data.

\section{Summary and Discussion}
\label{sec:summary}

The purpose of this paper is twofold: we discuss some new numerical features that we have successfully adopted in our study of critical collapse of gravitational waves, and we report on new results that complement the independent findings of  \AE\ and \cite{HilWB17,LedK21,FerRABH22} and that support our emerging understanding of these critical phenomena (see also \cite{Bauetal23}).

In terms of numerical features, we emphasize the importance of avoiding numerical error in the evaluation of the coefficients $A_\ell$, $B_\ell$, and $C_\ell$ (and their derivatives) that appear in the construction of Teukolsky wave initial data (see Sect.~\ref{sec:indata}).   We also discuss using both scalar invariants $\I$ and $\J$ of the Weyl curvature tensor as diagnostics of the spacetime geometry (Sect.~\ref{sec:diagnostics}).  Most importantly, however, we demonstrate the dramatic improvements that result from using a shock-avoiding slicing condition (see \cite{Alc97}) rather than the much more common 1+log slicing condition (Sect.~\ref{sec:slicing}; see also \cite{JimVA21}).  

While our fine-tuning to the onset of black-hole formation is more modest than in the simulations of \AE\ and \cite{HilWB17,LedK21,FerRABH22}, we have been able to identify, at least qualitatively, several features of the threshold solutions for both quadrupolar ($\ell = 2$) and hexadecapolar ($\ell = 4$) initial data.   For quadrupolar data we crudely estimate the critical exponent to be about $\gamma \simeq 0.25$, which is somewhat smaller than that reported by \cite{LedK21} for their (positive amplitude) Teukolsky waves, as well as that reported by \AE\ (even though neither \cite{LedK21} nor \AE\ adopt time-symmetric initial data).  

For both quadrupolar and hexadecapolar waves we find that the threshold solutions are consistent with featuring an approximate DSS with an accumulation event at the center.  For the quadrupolar data the period of this DSS is approximately $\Delta \simeq 0.5$, which is consistent with the value reported by \cite{AbrE94}, while for hexadecapolar data the period is significantly shorter, $\Delta \simeq 0.1$.   This difference in periodicity for different multipoles is consistent with that reported by \cite{PerB21} for the critical gravitational collapse of dipolar and quadrupolar electromagnetic waves, and demonstrates the absence of a universal critical solution: different multipoles lead to different threshold solutions.  

A further qualitative difference of the threshold solution for the hexadecapolar gravitational-wave data from that for quadrupolar data is that the periodic curvature peaks appear as pairs on the symmetry axis, on opposite sides from the center, similar to the observations of \cite{HilWB17,LedK21,FerRABH22}.  At least for the time-symmetric Teukolsky waves considered here, we believe that these peaks are features of a single self-similar solution with an accumulation event at the center, reminiscent of similar observations for quadrupolar electromagnetic waves \cite{PerB21}. It remains to be investigated what would happen for families of initial data that do not have a reflection symmetry, and so contain both even and odd $l$ spherical harmonics.

Contrary to our longstanding expectation, our results together with those of \cite{HilWB17,LedK21,FerRABH22} suggest that there does not exist a universal critical solution for the collapse of vacuum gravitational waves.  However, it appears that specific families of initial data may lead to threshold solutions that feature an at least approximately DSS region with an accumulation point at the center. Determining just how accurate this DSS is, and whether it holds up under better fine-tuning, will require future simulations with higher numerical accuracy.   In the meantime, we will discuss the implications of our findings, combined with those of \cite{HilWB17,LedK21,FerRABH22}, on our understanding of critical phenomena in the collapse of vacuum gravitational waves in a forthcoming joint article (see \cite{Bauetal23}).   

\acknowledgements

This study was supported by the Oberwolfach Research Fellows  program at the Mathematisches Forschungsinstitut Oberwolfach in 2022; we greatly appreciate the institute's and its staff's hospitality and support during our stay.  This work was also supported in part by National Science Foundation (NSF) grant PHY-2010394 to Bowdoin College, as well as by the FCT (Portugal) IF Programs IF/00577/2015 and PTDC/MAT-APL/30043/2017 and Project No.\ UIDB/00099/2020.  Numerical simulations were performed on the Bowdoin Computational Grid.

\begin{appendix}
\section{Linear gravitational waves}
\label{sec:appA}

In this appendix we list, as a reference, all expressions necessary to construct the linear gravitational waves in TT gauge that we use as a starting point for our initial data.  We adopt the convention and notation of Rinne \cite{Rin09}, who generalized Teukolsky's analysis for quadrupolar waves (see \cite{Teu82}) to higher multipole moments.  

We consider even-parity waves, for which the spacetime line element can be written as
\begin{align} \label{rinne_metric}
    ds^2 = & -dt^2 + (1 + A_\ell \hat Y^{\ell m}) dr^2 + 2 B_\ell \hat Y^{\ell m}_\theta r dr d\theta \nonumber \\
    & + 2 B_\ell \hat Y^{\ell m}_\varphi r \sin\theta dr d \varphi
    \nonumber \\
    & + (1 - A_\ell \hat Y^{\ell m} /2 + C_\ell \hat Y^{\ell m}_{\theta\theta}) \, r^2 d\theta^2 \nonumber \\
    & + 2 C_\ell \hat Y^{\ell m}_{\theta \varphi} r^2 \sin \theta d\theta d\varphi \nonumber \\
    & + (1 - A_\ell \hat Y^{\ell m} /2 + C_\ell \hat Y^{\ell m}_{\theta\theta} ) \, r^2 \sin^2 \theta d \varphi^2
\end{align}
(see \eq~4 in \cite{Rin09}).  In (\ref{rinne_metric}) the functions $A_\ell$, $B_\ell$, and $C_\ell$ (see \eqs~\ref{app:radial_2} and \ref{app:radial_4} below) depend on time $t$ and radius $r$ in the combinations $x \equiv r \pm t$ corresponding to ingoing and outgoing waves, and are computed from a seed function $F = F(x)$ (e.g.~\ref{seed}) and its derivatives, while the functions $\hat Y^{\ell m}$, $\hat Y^{\ell m}_\theta$, $\hat Y^{\ell m}_\varphi$, $\hat Y^{\ell m}_{\theta\theta}$, and $\hat Y^{\ell m}_{\theta\varphi}$ (see \ref{app:angular_2} and \ref{app:angular_4} below) depend on the angles $\theta$ and $\varphi$.  We specialize to axisymmetric solutions with $m = 0$, for which the functions $\hat Y^{\ell 0}_\varphi$ and $\hat Y^{\ell 0}_{\theta\varphi}$ vanish identically.  The specific expressions for $\ell = 2$ and $\ell = 4$, which we list below for convenience, are taken from  Appendix A of \cite{Rin09}.\footnote{We note a small typo in Appendix A of \cite{Rin09}, where the last entry for $\hat Y_{\theta\theta}$ for $\ell = 3$ should be $15 \cos \theta \sin^2 \theta$ rather than $15 \cos \theta \sin \theta$.}

\subsection{Quadrupolar waves: $\ell = 2$}
\label{sec:appA_l=2}

For quadrupolar waves with $\ell = 2$ the functions $A_2$, $B_2$, and $C_2$ are given by
\begin{subequations} \label{app:radial_2}
\begin{align}
    A_2 & = 24 \left( - \frac{F^{(2)}}{r^3} + \frac{3 F^{(1)}}{r^4} - \frac{3 F}{r^5} \right), \\
    B_2 & = 4 \left( - \frac{F^{(3)}}{r^2} + \frac{3 F^{(2)}}{r^3} - \frac{6 F^{(1)}}{r^4} + \frac{6 F}{r^5} \right), \\
    C_2 & = 2 \left( - \frac{F^{(4)}}{r} + \frac{2 F^{(3)}}{r^2} - \frac{3 F^{(2)}}{r^3} + \frac{3 F^{(1)}}{r^4} - \frac{3 F}{r^5} \right),
\end{align}
\end{subequations}
where $F^{(n)}(x) \equiv d^n F(x) / dx^n$ denotes the $n$-th derivative of the seed function $F$ with respect to its argument.  The non-vanishing angular functions are
\begin{subequations}  \label{app:angular_2}
    \begin{align}
        \hat Y^{20}  & = 2 - 3 \sin^2 \theta, \\
        \hat Y^{20}_\theta & = - 6 \cos \theta \sin \theta, \\
        \hat Y^{20}_{\theta\theta} & = 3 \sin^2 \theta.
    \end{align}
\end{subequations}

\subsection{Hexadecapolar waves: $\ell = 4$}
\label{sec:appA_l=4}

For hexadecapolar waves with $\ell = 4$, the functions $A_4$, $B_4$, and $C_4$ are given by
\begin{widetext}
\begin{subequations}  \label{app:radial_4}
\begin{align}
    A_4  & = 360 \left( - \frac{F^{(4)}}{r^3} + \frac{10 F^{(3)}}{r^4} - \frac{45 F^{(2)}}{r^5} 
    + \frac{105 F^{(1)}}{r^6} - \frac{105 F}{r^7} \right), \\
    B_4  & = 18 \left( - \frac{F^{(5)}}{r^2} + \frac{10F^{(4)}}{r^3} - \frac{55 F^{(3)}}{r^4} + \frac{195 F^{(2)}}{r^5} 
    - \frac{420 F^{(1)}}{r^6} + \frac{420 F}{r^7} \right), \\
    C_4  & = 2 \left( - \frac{F^{(6)}}{r} + \frac{9 F^{(5)}}{r^2} - \frac{45F^{(4)}}{r^3} + \frac{150 F^{(3)}}{r^4} - \frac{360 F^{(2)}}{r^5} + \frac{630 F^{(1)}}{r^6} - \frac{630 F}{r^7} \right),
\end{align}
\end{subequations}
\end{widetext}
while the non-vanishing angular functions are 
\begin{subequations}  \label{app:angular_4}
    \begin{align}
        \hat Y^{40} & = 35 \sin^4 \theta - 40 \sin^2 \theta + 8, \\
        \hat Y^{40}_\theta & = 20 \cos \theta \sin \theta \, (7 \sin^2 \theta - 4), \\
        \hat Y^{40}_{\theta\theta} & = 30 \sin^2 \theta \, (6 - 7 \sin^2 \theta).
    \end{align}
\end{subequations}

\end{appendix}

%


\end{document}